\documentstyle{article}

\author{Olavo, L. S. F. and Figueiredo, A. D.\\
Universidade de Brasilia - UnB\\
Departamento de Fisica - CEP 70910-900\\
Brasilia, D.F. - Brazil}
\title{Quantum Mechanics as a Classical Theory \\
VII: The Classical Spin Eigenfunctions
}

\begin{document}

\maketitle
\begin{abstract}
In this continuation paper the Schr\"odinger equation for the half-integral
spin eigenfunctions is obtained and solved. We show that all the properties
already derived using the Heisemberg matrix calculation and Pauli's matrices
are also obtained in the realm of these analytical functions. We also show
that Einstein-Bose condensation for fermions is expected. We then conclude
this series of two papers on the concept of classical spin.
\end{abstract}

\section{Introduction}

In paper VI (hereafter VI) of this series we developed a method by means of
which it was possible to derive the classical coordinate-momentum
representation for the behavior of particles with half-integral spin.

It was shown there that the matrix representation of this calculation is
readily obtained when we pass from the active to the passive views and
change the Poisson bracket to its similar matrix commutator. Then all matrix
mechanics related to the spin property was derived by straightforward
calculations.

We will now make use of the fact that our classical calculations give us the
coordinate-momentum representation of the spin property which is suitable to
be quantized---in the Schr\"odinger or Dirac's sense---to derive the
Schr\"odinger equation for the spin eigenfunction. This will be done in the
second section of this paper. This will show that Heisemberg's matrix
calculus and Schr\"odinger's analytical equation give identical results.

In the third section the Schr\"odinger equation for the spin will be solved
and we will show that all the properties derived by means of the algebraic
calculations are also obtainable with analytical functions.

We will devote the fourth section to show quantitativelly that fermionic
Einstein-Bose condensation is expected by the present theory.

In the last section we will make our conclusions.

\section{The Spin Schr\"odinger Equation}

We are now in position to derive the Schr\"odinger equation that will enable
us to obtain the half-integral spin eigenfunctions.

The two functions
\begin{equation}
\label{0}S^2\mbox{ and }S_3
\end{equation}
have to be written as operators in a quantization procedure. The spin
eigenfunction will be the function that makes these operators diagonal.

The quantization procedure has to be undertaken with much care since the
function $S^2$ has terms with products of position and momentum operators
which do not commute. To see this one needs only to look at this function
written in the coordinate-momentum representation
\begin{equation}
\label{1}S^2=\frac 1{16}\left[ \frac \alpha \beta \left( x^2+y^2\right)
^2+\frac \beta \alpha \left( p_x^2+p_y^2\right) ^2+2\left( x^2+y^2\right)
\left( p_x^2+p_y^2\right) \right] .
\end{equation}

This task is greatly simplified if we note that we might write
\begin{equation}
\label{2}S^2=\frac 14S_0^2,
\end{equation}
where
\begin{equation}
\label{3}S_0=\frac 12\left[ \sqrt{\frac \alpha \beta }\left( x^2+y^2\right)
+ \sqrt{\frac \beta \alpha }\left( p_x^2+p_y^2\right) \right] .
\end{equation}

This states the very difference from this problem to the one usually found
in textbooks on the solution of orbital angular momentum Schr\"odinger
equations. While in the later case one has only the operator ${\bf L}^2$
with the operator $L$, its square root, unknown, in the present approach
both are known by principle and their relation is defined by equation (\ref
{2}). This means that we will have only to find the function that makes the
operators $S_0$ and $S_3$ diagonal to automatically make $S^2$ also diagonal.

The next step is to transform the coordinate-momentum representation of our
problem into an operator representation.

We begin with equation (\ref{1}) and quantize the function $S_0$ in the
usual way, giving
\begin{equation}
\label{a1}\widehat{S}_0=\frac 12\left[ \sqrt{\frac \alpha \beta }\left(
\widehat{x}^2+\widehat{y}^2\right) +\sqrt{\frac \beta \alpha }\left(
\widehat{p}_x^2+\widehat{p}_y^2\right) \right]
\end{equation}
and also
\begin{equation}
\label{a2}\widehat{S}^2=\frac 1{16\gamma ^2}\left[ \left( \widehat{p}_x^2+
\widehat{p}_y^2\right) +\gamma ^2\left( \widehat{x}^2+\widehat{y}^2\right)
\right] ^2,
\end{equation}
where
\begin{equation}
\label{a3}\gamma =\sqrt{\frac \alpha \beta }.
\end{equation}

We might develop the product represented in expression (\ref{a2}) as a
repeated application of a differential operator to get%
$$
\widehat{S}^2=\frac 1{16\gamma ^2}\left\{ \left[ \left( \widehat{p}_x^2+
\widehat{p}_y^2\right) ^2+\gamma ^4\left( \widehat{x}^2+\widehat{y}^2\right)
^2+2\gamma ^2\left( \widehat{x}^2+\widehat{y}^2\right) \left( \widehat{p}%
_x^2+\widehat{p}_y^2\right) \right] -\right.
$$
\begin{equation}
\label{a4}\left. -4i\hbar \gamma ^2\left( \widehat{x}\widehat{p}_x+\widehat{y%
}\widehat{p}_y\right) -4\gamma ^2\hbar ^2\right\}
\end{equation}
where the crossed terms in momentum-coordinates were treated using Dirac's
symmetrization procedure according to which
\begin{equation}
\label{a5}x^2p_x^2\rightarrow \frac 12\left( \widehat{x}^2\widehat{p}_x^2+
\widehat{p}_x^2\widehat{x}^2\right) =\widehat{x}^2\widehat{p}_x^2-2i\hbar
\widehat{x}\widehat{p}_x^2-\hbar ^2,
\end{equation}
or simply developing the squared operator applied upon some function. The
first four terms in expression (\ref{a4}) represent $\widehat{S}_0^2$ which
means that we might write this expression as
\begin{equation}
\label{13}\widehat{S}^2=\frac 14\widehat{S}_0^2-\frac{\hbar ^2}4
\end{equation}
as our final operator $\widehat{S}^2$.

Looking at expressions (\ref{13}) we see that, if equation
\begin{equation}
\label{14}\widehat{S}_0\psi =\hbar \lambda \psi
\end{equation}
is satisfied, then the equation
\begin{equation}
\label{15}\widehat{S}^2\psi =\left( \frac 14\widehat{S}_0^2-\frac{\hbar ^2}%
4\right) \psi =\hbar ^2\left( \frac{\lambda ^2-1}4\right) \psi
\end{equation}
is automatically satisfied and we have a relation between the eigenvalues of
(\ref{14}) and (\ref{15}).

Equation (\ref{14}) might be written, in a differential form and in
rectangular coordinates, as
\begin{equation}
\label{16}\frac 12\left[ -\hbar ^2\sqrt{\frac \beta \alpha }\left( \frac{%
\partial ^2}{\partial x^2}+\frac{\partial ^2}{\partial y^2}\right) \psi +
\sqrt{\frac \alpha \beta }\left( x^2+y^2\right) \psi \right] =\hbar \lambda
\psi .
\end{equation}
We might now introduce polar coordinates
\begin{equation}
\label{17}x=r\cos \theta \mbox{ ; }y=r\sin \theta
\end{equation}
and
\begin{equation}
\label{18}\rho =\left( \frac \alpha {\beta \hbar ^2}\right) ^{1/4}r
\end{equation}
to find equation
\begin{equation}
\label{19}-\frac 1\rho \frac \partial {\partial \rho }\left[ \rho \frac{%
\partial \psi }{\partial \rho }\right] +\left( \rho ^2-2\lambda -\frac
1{\rho ^2}\frac{\partial ^2}{\partial \theta ^2}\right) \psi =0
\end{equation}
together with the equation for $S_3$ given by
\begin{equation}
\label{19a}\frac{-i\hbar }2\frac{\partial \psi }{\partial \theta }=m\psi .
\end{equation}

Equation (\ref{19a}) might be solved by putting
\begin{equation}
\label{20}\psi \left( \rho ,\theta \right) =e^{i2m\theta /\hbar }R(\rho )
\end{equation}
to get for equation (\ref{19})
\begin{equation}
\label{21}-\frac 1\rho \frac \partial {\partial \rho }\left[ \rho \frac{%
\partial R}{\partial \rho }\right] +\left( \rho ^2-2\lambda +\frac{4m^2}{%
\rho ^2}\right) R=0.
\end{equation}

This equation will be solved in the next section. However, it is noteworthy
that, for the wave function to be a continuous function of $\theta $, when a
rotation by an angle of $2k\pi $ radians is performed, we must have only
half-integral values for $m$.

\section{Solution of the Equation}

The solution of equation (\ref{21}) follows a standard method which we will
only briefly sketch here. We begin by looking at the asymptotic behavior of
function $R(\rho )$ for very large values of the variable $\rho $. For this
case we put
\begin{equation}
\label{23}R(\rho )\stackrel{\rho \rightarrow \infty }{\longrightarrow }%
e^{-k\rho ^2}
\end{equation}
to find
\begin{equation}
\label{23a0}k=+1/2
\end{equation}
as the only physically acceptable asymptotic behavior.

For small values of the variable $\rho $ we might try
\begin{equation}
\label{23a}R(\rho )\stackrel{\rho \rightarrow 0}{\longrightarrow }\rho ^s
\end{equation}
to find
\begin{equation}
\label{23b}s=\pm 2\left| m\right| ,\pm 2\left| m\right| -1,
\end{equation}
giving the four possible asymptotic behaviors
\begin{equation}
\label{24}R(\rho )\stackrel{\rho \rightarrow 0}{\longrightarrow }\rho ^{\pm
2\left| m\right| },\rho ^{\pm 2\left| m\right| -1},
\end{equation}
where the first two choices select the even series by letting $a_0$ free
(and making $a_1\equiv 0$) and the second two choices select the odd series
by letting $a_1$ free (and making $a_0\equiv 0$). We shall choose the plus
sign on both cases because the wave function has to be finite at the origin $%
\rho =0$. We then have
\begin{equation}
\label{24a}R(\rho )\stackrel{\rho \rightarrow 0}{\longrightarrow }\rho
^{2\left| m\right| },\rho ^{2\left| m\right| -1}.
\end{equation}

We next search for an expression for all values of $\rho $ using the series
expansion
\begin{equation}
\label{25}R(\rho )=\rho ^se^{-\frac 12\rho ^2}\sum_na_n\rho ^n.
\end{equation}
Substitution of expression (\ref{25}) into equation (\ref{21}) gives the
following equation
\begin{equation}
\label{26}\sum_n\rho ^{-2}\left\{ \left[ \left( s+n\right) ^2-4\left|
m\right| ^2)\right] a_n\rho ^n-2\left[ \lambda -(1+s+n)\right] a_n\rho
^{n+2}\right\} =0,
\end{equation}
which gives, for the coefficients, the recursion relation
\begin{equation}
\label{27}a_{n+2}=\frac{2\left[ (1+s+n)-\lambda \right] }{\left(
n+s+2\right) ^2-4\left| m\right| ^2}a_n.
\end{equation}
At this point it is easy to see that both choices of $s$ in equation (\ref
{24a}) will give exactly the same recursion relation and so, also the same
series---direct substitution of $s=2\left| m\right| $ with $n$ even and
substitution of $s=2\left| m\right| -1$ with $n$ odd shows this. We then
will work only with the choice
\begin{equation}
\label{27a}s=2\left| m\right| .
\end{equation}

If the series in expression (\ref{25}) does not terminate, its asymptotic
behavior when $n\rightarrow \infty $ is
\begin{equation}
\label{28}\frac{a_n}{a_{n-2}}\rightarrow \frac 2n,
\end{equation}
which is the same as the term $\rho ^ne^{\rho ^2}$ and is not acceptable as
a physical asymptotic behavior. Then the series shall terminate; this is
accomplished by making the choice
\begin{equation}
\label{29}\lambda _N=(1+2\left| m\right| +N),
\end{equation}
for some value of $n=N$. Since $N$ must be a positive number, we
automatically find the relation
\begin{equation}
\label{29a}\left| m\right| \leq \frac{\lambda _N-1}2.
\end{equation}

The correct eigenfunction of the problem is given by
\begin{equation}
\label{30}R(\rho )=\rho ^{2\left| m\right| }e^{-\frac 12\rho
^2}\sum_{n=0}^Na_n\rho ^n\mbox{ }
\end{equation}
and is a solution of the system
\begin{equation}
\label{31}S_3\psi =\hbar m\psi \mbox{ ; }S^2\psi =\hbar ^2\left( \frac{%
\lambda _N-1}2\right) \left( \frac{\lambda _N+1}2\right) \psi .
\end{equation}
Comparing this last expression with
\begin{equation}
\label{32}S^2\psi =\hbar ^2\ell \left( \ell +1\right) \psi ,
\end{equation}
we find that
\begin{equation}
\label{33}\ell =\frac{\lambda _N-1}2.
\end{equation}
Equation (\ref{29a}) then means that
\begin{equation}
\label{34}\left| m\right| \leq \ell
\end{equation}
as expected.

The multiplicity of our functions might be calculated with the use of the
quantum number $\ell $. One might easily check that this multiplicity is
given by
\begin{equation}
\label{34a}2\ell +1=\lambda _N.
\end{equation}

The final eigenfunction of our problem might be written in the $(r,\theta )$
representation as
\begin{equation}
\label{35}\psi (r,\theta )=\left( \frac \alpha {\beta \hbar ^2}\right)
^{\left| m\right| /2}r^{2\left| m\right| }e^{2im\theta /\hbar }e^{-\frac
12\left( \alpha /\beta \hbar ^2\right) ^{1/2}r^2}\sum_{n=0}^Na_n\left( \frac
\alpha {\beta \hbar ^2}\right) ^{n/4}r^n,\mbox{ }
\end{equation}
where the coefficients $a_n$ are given by expression (\ref{27}) and the
ratio $\alpha /\beta $ is a structure constant used to identify the actual
particle we are interested in---as seen in expression (\ref{a1}). This
structure constant is necessary since our calculations were general and
reflect the behavior of any half-integer spin particle. It is then possible
to calculate quantities such as the radius of the particle in terms of this
constant by performing the integral
\begin{equation}
\label{35a}\overline{r}_{\ell ,\left| m\right| }(\alpha /\beta )=\int \psi
_{\ell ,\left| m\right| }^{\dagger }(r,\theta )r\psi _{\ell ,\left| m\right|
}(r,\theta )rdrd\theta .
\end{equation}
As an example we might find the radius of a half spin particle $(m=1/2,\ell
=1/2)$ using its density function
\begin{equation}
\label{be3}d(r)=\left| \psi \left( r,\theta \right) \right|
^2=N^2r^2e^{-\left( \frac \alpha {\beta \hbar }\right) ^{1/2}\ r^2},
\end{equation}
where $N$ is a normalization constant. This density is related with the
internal structure distribution of the half-integral spin particles and
shall not be interpreted as probability distributions---we might get the
mass distribution of the electron by multiplying the above expression by its
mass, for example.

The mean radius of this particle is
\begin{equation}
\label{be5}<r_{1/2,1/2}>=\frac{\int_0^\infty r^4e^{-\frac \alpha {\beta
\hbar }\ r^2}dr}{\int_0^\infty r^3e^{-\frac \alpha {\beta \hbar }\ r^2}dr}=
\frac{3\sqrt{\pi \hbar }}4\left( \frac \alpha \beta \right) ^{-1/4},
\end{equation}
where we notice that the bigger the structure constant, the smallest the
particle.

The first possible values of $\lambda $ are given in table I. In this table
we show the $\ell $ value related to the chosen $\lambda $ value and all
possible values of $\left| m\right| $. The values of the cutoff number $N$
and the multiplicity associated with $\ell $ is also shown. It is remarkable
that if we try to put odd values for variable $\lambda $, or else, integer
values for variable $\ell $, we cannot terminate the series in (\ref{35}),
since in this case the cutoff number $N$ is odd, which is not allowed by our
choice of the asymptotic behavior at the origin in expression (\ref{27a}).
Since the series does not terminate its asymptotic behavior for large values
of variable $\rho $ is no more given by expression (\ref{23}) and we must
reject these solutions.

We also plot the density distributions for the cases $(\ell =1/2,\left|
m\right| =1/2)$, $(\ell =7/2,\left| m\right| =7/2)$ and $(\ell =9/2,\left|
m\right| =3/2)$ in figures 1, 2 and 3 respectively.

Figure 1 shows that the $\ell =1/2$ spin particles structure is somewhat
like a ring (zero density at the origin) with maximum density at a radius $%
r_{\ell =1/2,\left| m\right| =1/2}$, depending on the
structure constant.

In figure 2 the same behavior of figure 1 is attained but we might see that
the distance of the maximum density from the origin is now $r_{\ell
=7/2,\left| m\right| =7/2}(\alpha /\beta )$ which is greater than the one
for the half spin case. We can visualize this as the increasing of a
`centrifugal' force giving the screening of the particle matter
distribution---we are considering the same value for the structure constant.

Figure 3 shows that the same ring structure will be present in all functions
(they depend on $r^{2\left| m\right| }$ which always makes the densities
tend to zero in the vicinities of the origin). For the cases where the
difference in the quantum numbers $\ell $ and $\left| m\right| $ are
different from zero we also find the appearance of nodes reflecting the
multiple ring structure of these particles. The number of these concentric
rings will be given by the expression $(\ell -\left| m\right| )+1$.

\section{Bose-Einstein Condensation}

In the previous paper (VI) we have shown that, if the parameter $\lambda $
has a lower bound, then the phenomenon of Bose-Einstein condensation will be
expected for some temperature $T_{cond}$. This is precisely what we have
found. In Table I we note that the parameter $\lambda $ has the value $%
\lambda _{\min }=2$ as a lower bound.

This implies that the fermion shall be supplied with at least the energy
\begin{equation}
\label{be10}E_{\min }=2\hbar \omega
\end{equation}
to continue to act as a fermion.

Because of this lower bound for the energy of the fermions we expect that,
at some value of the temperature, condensation takes place.

\section{Conclusion}

In these two papers we aimed at showing that: (1) the concept of spin is not
a particularity of the quantum mechanical formalism and might also be
represented in the realm of classical mechanics from where we can extract a
{\em model} for it (or a picture). (2) even in this case `space
quantization' might be obtained apart from a constant (phenomenologically
obtainable) which we relate to Planck's constant. (3) the concept of spin is
related to the symmetries generated by the Lie algebra associated with its
Lie Group (SU(2)); since this group is the same generated by classical
phase-space functions obeying the same Lie algebra with the product defined
as the Poisson bracket, there is no impossibility in deriving the concept of
classical spin. (4) this classical representation of the spin might be
`quantized' using traditional methods to derive a Schr\"odinger equation
which is an analytical representation for this quantity. The solution of
this equation, when squared, will give us information on the particle
internal structure, as for example its mass or charge distribution or its
mean radius, in terms of some characteristic constant related to each
particle and identifying it. (5) all expected quantum properties already
obtained by Heisemberg matrix calculations, using Pauli's matrices, are also
obtained with this method and this result reaffirms the formal identity
between Heisemberg's matrix calculus and Schr\"odinger's one. (6) spin is
not a characteristic of relativistic calculations although it is better
represented in the realm of this theory where Lorentz invariance might be
imposed. (7) fermionic Bose-Einstein condensation has a very intuitive
explanation by means of the minimum energy the fermions have to possess in
order to behave like fermions.

We hope that these calculations will help in clarifying some misconceptions
rather diffused in the literature about the classical {\it versus} quantum
status of half-spin particles and also about the possibility of an
analytical representation of the concept of half-spin.

Aside the epistemological aspects, we also hope that the possession of the
spin eigenfunctions will help the investigations in areas such as
superconductivity, Bose-Einstein condensation and many others.

\section{Acknowledgements}

The authors wish to thanks the Conselho Nacional de Desenvolvimento
Cient\'ifico e Tecnol\'ogico  (CNPq) for sponsoring this research.

{\normalsize
\begin{table}
 \begin{center}
  \begin{tabular}{|c|c|c|c|c|} \hline
{$\lambda$}&$\ell$&$\left|m\right|$&{\bf N}&{\bf Multiplicity}\\ \hline \hline
    2      & 1/2   &  1/2           &  0    &       2          \\ \hline
    3      &  1    &  1/2           &  1    &   $\infty$       \\ \hline
    4      & 3/2   &  1/2           &  2    &                  \\
           &       &  3/2           &  0    &       4          \\ \hline
    5      &  2    &  1/2           &  3    &   $\infty$       \\
           &       &  3/2           &  1    &   $\infty$       \\ \hline
    6      & 5/2   &  1/2           &  4    &                  \\
           &       &  3/2           &  2    &                  \\
           &       &  5/2           &  0    &       6          \\ \hline
    7      & 3     &  1/2           &  5    &   $\infty$       \\
           &       &  3/2           &  3    &   $\infty$        \\
           &       &  5/2           &  1    &   $\infty$       \\ \hline
    8      & 7/2   &  1/2           &  6    &                  \\
           &       &  3/2           &  4    &                   \\
           &       &  5/2           &  2    &                   \\
           &       &  7/2           &  0    &       8           \\ \hline
    9      & 4     &  1/2           &  7    &   $\infty$        \\
           &       &  3/2           &  5    &   $\infty$        \\
           &       &  5/2           &  3    &   $\infty$        \\
           &       &  7/2           &  1    &   $\infty$        \\ \hline
    10     & 9/2   &  1/2           &  8    &                   \\
           &       &  3/2           &  6    &                   \\
           &       &  5/2           &  4    &                   \\
           &       &  7/2           &  2    &                   \\
           &       &  9/2           &  0    &   10   \\  \hline \hline \hline
  \end{tabular}
\end{center}
 \caption{Values of $\lambda$ or $\ell$ in terms of $m$ and $N$.
          The multiplicity of each choice is also shown. \label{tabela1}}
\end{table}
}

\end{document}